\documentclass[pra,twocolumn,showpacs,preprintnumbers,superscriptaddress]{revtex4-1}
\usepackage{times}
\usepackage{bm}
\usepackage{graphicx}
\usepackage{amsbsy}
\usepackage{amsmath}
\usepackage{amsfonts}
\usepackage{amsthm}
\usepackage{xcolor}
\usepackage{empheq}

\begin{document}

\theoremstyle{plain}
\newtheorem{theorem}{Theorem}
\newtheorem{lemma}[theorem]{Lemma}
\newtheorem{corollary}[theorem]{Corollary}
\newtheorem{proposition}[theorem]{Proposition}\newtheorem{conjecture}[theorem]{Conjecture}
\theoremstyle{definition}
\newtheorem{definition}[theorem]{Definition}
\theoremstyle{remark}
\newtheorem*{remark}{Remark}
\newtheorem{example}{Example}
\title{Estimation of Entanglement Negativity of a Two-Qubit Quantum System With Two Measurements}
\author{Satyabrata Adhikari}
\email{satyabrata@dtu.ac.in} \affiliation{Delhi Technological
University, Shahbad Daulatpur, Main Bawana Road, Delhi-110042,
India}

\begin{abstract}
Numerous work had been done to quantify the entanglement of a
two-qubit quantum state, but it can be seen that previous works
were based on joint measurements on two copies or more than two
copies of a quantum state under consideration. In this work, we
show that a single copy and two measurements are enough to
estimate the entanglement quantifier like entanglement negativity
and concurrence. To achieve our aim, we establish a relationship
between the entanglement negativity and the minimum eigenvalue of
structural physical approximation of partial transpose of an
arbitrary two-qubit state. The  derived relation make possible to
estimate entanglement negativity experimentally by Hong-Ou-Mandel
interferometry with only two detectors. Also, we derive the upper
bound of the concurrence of an arbitrary two-qubit state and have
shown that the upper bound can be realized in experiment. We will
further show that the concurrence of (i) an arbitrary pure
two-qubit states and (ii) a particular class of mixed states,
namely, rank-2 quasi-distillable mixed states, can be exactly
estimated with two measurements.
\end{abstract}
 \maketitle

\section{Introduction}
Entanglement lies at the heart of quantum mechanics
\cite{einstein} and act as an essential ingredient in quantum
computing \cite{steane}, quantum communication \cite{gisin1} and
quantum cryptography \cite{gisin2}. The amount of entanglement in
a given entangled system is directly proportional to the
efficiency of given entangled state in doing some quantum
information processing task. This implies that if the system is
maximally entangled or even non-maximally entangled two-qubit
system then these entangled systems are able to perform better
than separable states in quantum information processing task.
Therefore, it is necessary to determine whether the generated
quantum state is entangled or not. This problem is known as
entanglement detection problem and it is studied by many authors
applying different techniques \cite{guhne, peres, horodecki1,
rossignoli, gao, lima, shahandeh, ganguly, sadhikari}. The
entanglement can also be detected by structural physical
approximation method (SPA) \cite{wang, kalev}. SPA is a physical
means by which positive maps can be
approximated by completely positive maps \cite{horodecki2}.\\
Like entanglement detection problem, entanglement quantification
is also equally important. To develop the quantum technology
\cite{amico} and perform quantum communication task \cite{gisin1},
it is important to know the exact amount of entanglement in an
entangled state. Large effort has been devoted to quantify the
amount of entanglement by defining different entanglement measures
such as entanglement of formation \cite{wootters}, entanglement
negativity \cite{vidal}, logarithmic negativity \cite{plenio},
relative entropy of entanglement \cite{vedral}. Entanglement of
formation is a function of a quantity called concurrence and same
optimal pure state decomposition can be used to calculate
concurrence and entanglement of formation
\cite{wootters,uhlmann1}. Multiple effort had been done for the estimation of
entanglement of two qubit system without quantum
state tomography (QST) \cite{bartkiewicz, bartkiewicz1, gray}. In a
similar fashion, numerous effort had been given to determine
concurrence for two qubit system without quantum state tomography
\cite{heydari,lee, bail, park}. Estimation of entanglement has been studied for multipartite higher dimensional
system also \cite{audenaert1, bavaresco, ghosh}. Walborn et.al. \cite{walborn} performed
an experiment to quantify entanglement using linear optics set up and
determine concurrence of pure state by performing only single
measurement on two copies. P. Horodecki \cite{phorodecki} provided
a protocol that uses collective measurements on at most eight
copies to determine the concurrence and negativity of a two qubit
system. L. H. Zhang et.al. \cite{zhang} presented protocols for
the direct measurement of the concurrence for two-photon
polarization entangled pure and mixed states without tomography.
Their protocols need two copies of the
state in each detection round.\\
It is clear from the above discussion that the methods proposed
earlier for the estimation of either entanglement negativity or
concurrence or both for a two-qubit system either require more
than a single copy of the quantum state under investigation or
require to estimate more than two parameters. As far as my
knowledge, we find that there does not exist any single method
that rely on single copy and two measurements for the estimation
of entanglement negativity and concurrence of a two-qubit system.
This motivate us to present a method that need one parameter
estimation and single copy of the quantum state under
investigation. In the present work, we will apply SPA of partial
transpose (PT) for the
estimation of entanglement negativity and concurrence for two-qubit system.\\
This paper is organized as follows: In Sec. II, we discuss about
the entanglement negativity and point out the difficulty in
realizing it in experiment. Also, we derive the lower bound of the
entanglement negativity using a mathematical inequality involving
trace of the product of two Hermitian matrices and product of the
eigenvalues of the same Hermitian matrices. In Sec. III, we
provide the exact value of the entanglement negativity in terms of
the minimum eigenvalue of SPA-PT of an arbitrary two-qubit mixed
state $\rho_{AB}$. In Sec. IV, we illustrate our results with
examples. In Sec. V, we show that it is possible to express the
exact value of the concurrence of the pure entangled state and
quasi-distillable mixed state in terms of the minimum eigenvalue
of SPA-PT of the two-qubit pure entangled state
$|\psi\rangle_{AB}$ and quasi-distillable mixed state
$\rho_{AB}^{quasi}$. In Sec. VI, we determine the upper bound of
the concurrence of an arbitrary two-qubit density matrix. In Sec.
VII, we summarize our result.
\section{Lower bound of entanglement negativity}Entanglement negativity
$N(\rho)$ for the density matrix $\rho$ is defined as
\cite{miranowicz,horst}
\begin{eqnarray}
N^{D}(\rho)= 2\sum_{i}max(0,-\nu_{i})
 \label{def}
\end{eqnarray}
where $\nu_{i}'s$ are the negative eigenvalues of the partial
transpose $\rho^{\Gamma}$ of the density matrix $\rho$.\\
Entanglement negativity was introduced by Vidal and Werner and
have shown that it is indeed an entanglement monotone
\cite{vidal}. The number of negative eigenvalues of
$\rho^{\Gamma}$ is at most two when $\rho$ denoting a two qubit
system \cite{rana}. A point to be noted is that, we restrict
ourselves in this work to two-qubit system. Thus we need to
determine only one negative eigenvalue of the partial transpose of
a two-qubit state provided the quantum state is entangled.
Although it is one of the important measure of entanglement for
bipartite as well as multi-partite system but it involves PT,
which is a positive but not completely positive map. This means
that it is very difficult to realize PT in the laboratory and
hence restrict the determination of entanglement negativity
experimentally. To overcome this difficulty, we will apply the
method of SPA of PT map. Due to this approximation, PT map reduces
to a completely positive map that can be realized in the
experiment. The experimental demonstration of SPA-PT for two qubit
photonic system using single-photon polarization qubits and linear
optical devices had been given by H-T Lim et.al. \cite{lim}. We
note that an important application of SPA-PT method had been
discussed recently in \cite{adhikari}, where it had been shown
that SPA-PT estimated optimal singlet fraction with only two measurements.\\
In order to estimate the exact value of $N(\rho)$ experimentally,
we first derive the lower bound of entanglement negativity in
terms of a quantity that can be realized in the experiment. To
proceed in this direction, let us first consider two subsystems
$A$ and $B$ described by the Hilbert spaces $H_{A}$ and $H_{B}$
respectively, which are the part of the composite system described
by the Hilbert space $H_{AB}$. Consider any two-qubit entangled
state $\rho_{AB}$ in the composite Hilbert space $H_{AB}$. Now our
task is to derive the lower bound of entanglement negativity
$N(\rho_{AB})$, which quantify the amount of entanglement between
two subsystems $A$ and $B$ and show that the lower bound can be
realized in experiment with only two
measurements. To achieve this task, we start with the statement of a lemma.\\
\textsl{Lemma\cite{lasserre}:-} For any two Hermitian $4\times 4$
matrices $F_{1}$ and $F_{2}$, the inequality given below holds
true
\begin{eqnarray}
\sum_{i=1}^{4}\lambda_{i}(F_{1})\lambda_{5-i}(F_{2})\leq
Tr(F_{1}F_{2})\leq
\sum_{i=1}^{4}\lambda_{i}(F_{1})\lambda_{i}(F_{2})
 \label{lemma}
\end{eqnarray}
where $\lambda_{i}(F_{1})$ and $\lambda_{i}(F_{2})$ denote the
eigenvalues of the matrices $F_{1}$ and $F_{2}$ respectively. The
eigenvalues are arranged in descending order i.e.
$\lambda_{1}(.)>\lambda_{2}(.)>\lambda_{3}(.)>\lambda_{4}(.)$.\\
We use the above stated lemma for $F_{1}=|\phi\rangle\langle\phi|$
and $F_{2}=\rho_{AB}^{T_{B}}$, where
$|\phi\rangle=\alpha|00\rangle+\beta|11\rangle$, $T_{B}$ denotes the partial transposition with respect to the subsystem B and $\rho_{AB}$ denoting the two-qubit density operator expressed in the computational basis as
\begin{eqnarray}
\rho_{AB}=
\begin{pmatrix}
  t_{11} & t_{12} & t_{13} & t_{14} \\
  t_{12}^{*} & t_{22} & t_{23} & t_{24} \\
  t_{13}^{*} & t_{23}^{*} & t_{33} & t_{34} \\
  t_{14}^{*} & t_{24}^{*} & t_{34}^{*} & t_{44}
\end{pmatrix}, \sum_{i=1}^{4}t_{ii}=1
\end{eqnarray}
where $(*)$ denotes the complex conjugate.\\\\
As a result, the left hand inequality of (\ref{lemma}) will become
\begin{eqnarray}
\lambda_{4}(\rho_{AB}^{T_{B}})\leq
Tr(|\phi\rangle\langle\phi|\rho_{AB}^{T_{B}})
 \label{lhineq}
\end{eqnarray}
The amount of entanglement contained in $\rho_{AB}$ can be
determined by entanglement negativity $N(\rho_{AB})$ defined in
(\ref{def}). It is a positive real number and is given by
\begin{eqnarray}
N(\rho_{AB})=-2\lambda_{4}(\rho_{AB}^{T_{B}})
 \label{entneg}
\end{eqnarray}
Inserting (\ref{entneg}) in (\ref{lhineq}), we have
\begin{eqnarray}
-\frac{N(\rho_{AB})}{2}\leq
Tr(|\phi\rangle\langle\phi|\rho_{AB}^{T_{B}})
 \label{lhineq1}
\end{eqnarray}
Note that the value of the R.H.S of inequality (\ref{lhineq1})
cannot be determined experimentally because partial transposition
operation is not a physical operation. To obtain the value of
R.H.S of inequality (\ref{lhineq1}) experimentally, we use
structural physical approximation method to approximate the
non-physical partial transposition operation by a completely
positive operation. Let the structural physical approximation of
$\rho_{AB}^{T_{B}}$ be $\widetilde{\rho_{AB}}$ and it is given by
\begin{eqnarray}
\widetilde{\rho_{AB}}&=&[\frac{1}{3}(I\otimes\widetilde{T})+\frac{2}{3}(\widetilde{\Theta}\otimes D)]\rho_{12}\nonumber\\&=&
\begin{pmatrix}
  E_{11} & E_{12} & E_{13} & E_{14} \\
  E_{12}^{*} & E_{22} & E_{23} & E_{24} \\
  E_{13}^{*} & E_{23}^{*} & E_{33} & E_{34} \\
  E_{14}^{*} & E_{24}^{*} & E_{34}^{*} & E_{44}
\end{pmatrix}
\label{spa1}
\end{eqnarray}
where
\begin{eqnarray}
&&E_{11}=\frac{1}{9}(2+t_{11}),E_{12}=\frac{1}{9}(-it_{12}+t_{12}^{*}),\nonumber\\&&
E_{13}=\frac{1}{9}(t_{13}-i(t_{13}^{*}+t_{24}^{*})),
E_{14}=\frac{1}{9}(-it_{14}+t_{23}),\nonumber\\&&
E_{22}=\frac{1}{9}(2+t_{22}),E_{23}=\frac{1}{9}(t_{14}+it_{23}),\nonumber\\&&
E_{24}=\frac{-i}{9}(t_{13}^{*}+t_{24}^{*}),E_{33}=\frac{1}{9}(2+t_{33}),\nonumber\\&&
E_{34}=\frac{1}{9}(-it_{34}+t_{34}^{*}),E_{44}=\frac{1}{9}(2+t_{44})
\label{spa2a}
\end{eqnarray}
$I\otimes\tilde{T}$ and
$\tilde{\Theta}\otimes D$ are local operations and are completely
positive operators. The SPAed transpose $\tilde{T}$ for a density operator $\rho$ is given by
\begin{eqnarray}
\tilde{T}(\rho)=\sum_{k=1}^{4}Tr(M_{k}\rho)|s_{k}\rangle\langle s_{k}|
\label{spatrans}
\end{eqnarray}
where $\{M_{k}=\frac{1}{2}|s_{k}^{*}\rangle\langle s_{k}^{*}|\}_{k=1}^{4}$ is a complete measurement and\\
$|s_{1}^{*}\rangle=\frac{1}{\sqrt{1+|b_{1}|^{2}}}(|0\rangle+b_{1}^{*}|1\rangle)$,
$|s_{2}^{*}\rangle=\frac{1}{\sqrt{1+|b_{1}|^{2}}}(|0\rangle-b_{1}^{*}|1\rangle)$,
$|s_{3}^{*}\rangle=\frac{1}{\sqrt{1+|b_{2}|^{2}}}(|0\rangle+b_{2}^{*}|1\rangle)$,
$|s_{4}^{*}\rangle=\frac{1}{\sqrt{1+|b_{2}|^{2}}}(|0\rangle-b_{2}^{*}|1\rangle)$,\\ $b_{1}=\frac{ie^{\frac{2i\pi}{3}}}{i+e^{\frac{-2i\pi}{3}}}$,
$b_{2}=\frac{ie^{\frac{2i\pi}{3}}}{i-e^{\frac{-2i\pi}{3}}}$.\\
The other local operators $\tilde{\Theta}$ and $D$ are defined as
 \begin{eqnarray}
\tilde{\Theta}(.)=\sigma_{y}\tilde{T}(.)\sigma_{y}
\label{spatheta}
\end{eqnarray}
\begin{eqnarray}
D(.)=\frac{1}{4}\sum_{i=0,x,y,z}\sigma_{i}(.)\sigma_{i}
\label{spaD}
\end{eqnarray}
where $\sigma_{0}=I$ and $\sigma_{x}$,$\sigma_{y}$,$\sigma_{z}$ denote the Pauli matrices.\\
The quantum circuit for the realization of SPA-PT operation has been designed in
\cite{kalev}.\\
The relation between $\rho_{AB}^{T_{B}}$ and $\widetilde{\rho_{AB}}$ is given
by \cite{adhikari}
\begin{eqnarray}
Tr(|\phi\rangle\langle\phi|\rho_{AB}^{T_{B}})=9Tr[|\phi\rangle\langle\phi|\widetilde{\rho_{AB}}]-2
 \label{spa1}
\end{eqnarray}
Using (\ref{lhineq1}) and (\ref{spa1}), we get
\begin{eqnarray}
Tr[|\phi\rangle\langle\phi|\widetilde{\rho_{AB}}]\geq\frac{1}{18}[4-N(\rho_{AB})]
\label{lhineq2}
\end{eqnarray}
Let $\mu_{min}$ be the minimum eigenvalue of
$\widetilde{\rho_{AB}}$ then the eigenvalue equation is given by
\begin{eqnarray}
\widetilde{\rho_{AB}}|\phi\rangle=
\mu_{min}|\phi\rangle\label{relation}
\end{eqnarray}
Using the eigenvalue equation (\ref{relation}) in (\ref{lhineq2}),
we get
\begin{eqnarray}
N(\rho_{AB})\geq 4-18\mu_{min} \label{lowerbound}
\end{eqnarray}
The minimum eigenvalue $\mu_{min}$ of the quantum state
$\widetilde{\rho_{AB}}$ can be determined by the formula \cite{adhikari}
\begin{eqnarray}
\mu_{min} = \frac{15}{8}F_{avg}(\widetilde{\textit{\textbf{W}}},\widetilde{\rho_{AB}})-\frac{47}{72}\label{mineig}
\end{eqnarray}
where $F_{avg}(\widetilde{\textit{\textbf{W}}},\widetilde{\rho_{AB}})$
and $\widetilde{\textit{\textbf{W}}}$ denoting the average
fidelity and approximated entanglement witness operator
respectively. The approximated entanglement witness operator can
be expressed in terms of entanglement witness operator
$\textit{\textbf{W}}=|\phi\rangle\langle\phi|-\frac{2}{9}I$ as
\cite{adhikari}
\begin{eqnarray}
\widetilde{\textit{\textbf{W}}}=\frac{2}{9}\textit{\textbf{W}}+\frac{7}{36}I,
\label{aew1}
\end{eqnarray}
The average fidelity $F_{avg}(\widetilde{\textit{\textbf{W}}},\widetilde{\rho_{AB}})$ can be
determined experimentally with only two measurement by using Hong-Ou-Mandel interferometry
\cite{kwong,adhikari} and hence the minimum eigenvalue $\mu_{min}$. Thus we can determine the value of lower
bound of entanglement negativity experimentally with only two
measurement.\\
\section{Determination of exact value of entanglement
negativity} We have already obtained the analytical lower bound of
entanglement negativity and it is given by (\ref{lowerbound}).
Also we have shown that the analytic lower bound can be achieved
experimentally. Now the problem is that the inequality
(\ref{lowerbound}) only tells us that the entanglement negativity
can take value greater than the bound obtained but it does not
determine the actual amount of entanglement in an arbitrary
two-qubit state. We are now interested in obtaining the actual
value of the entanglement negativity contained in an arbitrary
two-qubit state. Note that the quantity $4-18\mu_{min}$ is less
than or equal to $N(\rho_{AB})$. This suggest that if we add a
positive quantity $Q$ to the quantity $4-18\mu_{min}$ then it may
be equal to $N(\rho_{AB})$. By adding a positive quantity $Q$ to
the R.H.S of (\ref{lowerbound}), we get
\begin{eqnarray}
N(\rho_{AB})= 4-18\mu_{min}+ Q, Q>0
 \label{q1}
\end{eqnarray}
To search for the quantity $Q$, we keep in mind the following
facts: (i) the inequality $\mu_{min}<\frac{2}{9}$ holds for all
entangled state $\rho_{AB}$ \cite{horodecki2} and (ii)
$Tr[(I-|\phi\rangle\langle\phi|)\widetilde{\rho_{AB}}]=1-\mu_{min}>0$.
Using these two facts, we can always choose
$Q=(\frac{2}{9}-\mu_{min})Tr[(I-|\phi\rangle\langle\phi|)\widetilde{\rho_{AB}}]$.
With this choice of $Q$, (\ref{q1}) can be re-written as
\begin{eqnarray}
N(\rho_{AB})&=& 4-18\mu_{min}+
(\frac{2}{9}-\mu_{min})(1-\mu_{min})
\nonumber\\&=&(\frac{2}{9}-\mu_{min})(19-\mu_{min}),
\nonumber\\&&\frac{1}{6}\leq \mu_{min} < \frac{2}{9}\label{en1}
\end{eqnarray}
Here, we observe that $N(\rho_{AB})$ given in (\ref{en1}) is not
normalized. The normalized $N(\rho_{AB})$ is then given by
\begin{eqnarray}
N^{N}(\rho_{AB})= K(\frac{2}{9}-\mu_{min})(19-\mu_{min}),
\frac{1}{6}\leq \mu_{min} < \frac{2}{9} \label{en}
\end{eqnarray}
Here $K$ is a normalization constant. $K$ can be determined by
using the fact that $\mu_{min}=\frac{1}{6}$ for maximally
entangled state. Thus, the normalized entanglement negativity is
given by
\begin{eqnarray}
N^{N}(\rho_{AB})&=&
\frac{108}{113}(\frac{2}{9}-\mu_{min})(19-\mu_{min}),\nonumber\\&&
\frac{1}{6}\leq \mu_{min} < \frac{2}{9} \label{en2}
\end{eqnarray}
Since $N^{N}(\rho_{AB})$ expressed in terms of $\mu_{min}$ so the
normalized entanglement negativity can be determined
experimentally with two measurement using Hong-Ou-Mandel
interferometry.\\
Thus we have presented a entanglement estimation protocol in which we need only single copy of the quantum state
and no requirement of QST. This result contradict the fact that it is impossible to detect the value of the entanglement with only single-copy measurements without QST \cite{lu, carmeli}. The above contradiction can be explained by observing the fact that there does not exist any quantum operation that can achieve non-physical operation such as partial transpose map with unit fidelity in an experiment. Fidelity measures how much close the approximate quantum operation to the actual impossible operation \cite{bae}. By seeing this fact, it is necessary to have approximate quantum operation that can approximate partial transpose for the possible realization in experiment.  SPA is such an approximation of PT operation that can be realized in experiment with fidelity less than unity. Since our protocol for the estimation of entanglement based on SPA-PT so the estimation is an approximate estimation. Further, the parameter needed for the estimation of entanglement is the minimum eigenvalue of the SPA of the given state
and this parameter is related with the average fidelity. So, the minimum eigenvalue can be estimated approximately and
hence the negativity.\\
\section{Examples} We now illustrate with examples that the
derived expression of normalized entanglement negativity given in
(\ref{en2}) is indeed correct and its correctness can be shown by
matching it with the expression of entanglement negativity
obtained via definition (\ref{def}).\\
\textbf{Example-1:}Let us first consider a pure entangled state
described by density matrix
\begin{eqnarray}
\rho_{M}&=&
M|01\rangle\langle01|+\sqrt{M(1-M)}(|01\rangle\langle10|+|10\rangle\langle01|)
\nonumber\\&&+(1-M)|10\rangle\langle10| \label{pureent}
\end{eqnarray}
The entanglement negativity using the definition (\ref{def}) is
given by
\begin{eqnarray}
N^{D}(\rho_{M})=2\sqrt{M(1-M)} \label{entnegdef1}
\end{eqnarray}
The SPA-PT of $\rho_{M}$ is given by
\begin{eqnarray}
\widetilde{\rho_{M}}=
\begin{pmatrix}
  \frac{2}{9} & 0 & 0 & a \\
  0 & \frac{2+M}{9} & ia & 0 \\
  0 & -ia & \frac{3-M}{9} & 0 \\
   a  & 0 & 0 & \frac{2}{9}
\end{pmatrix}
\label{spapt1}
\end{eqnarray}
where $a=\frac{\sqrt{M(1-M)}}{9}$.\\
The minimum eigenvalue of $\widetilde{\rho_{M}}$ is given by
\begin{eqnarray}
\mu_{min}&=&\frac{2}{9}-\frac{\sqrt{M(1-M)}}{9}\nonumber\\&=&
\frac{15}{8}F_{avg}(\widetilde{\textit{\textbf{W}}},\widetilde{\rho_{M}})-\frac{47}{72}
 \label{mineig1}
\end{eqnarray}
where
$F_{avg}(\widetilde{\textit{\textbf{W}}},\widetilde{\rho_{M}})$
denoting the average fidelity between two quantum state $\widetilde{\textit{\textbf{W}}}$
and $\widetilde{\rho_{M}}$.\\
The value of $\mu_{min}$ given by the second equality in
(\ref{mineig1}) can be estimated experimentally by Hong-Ou-Mandel
interferometer set up with only two detectors
\cite{adhikari,kwong}. Substituting
$\mu_{min}=\frac{2}{9}-\frac{\sqrt{M(1-M)}}{9}$ in (\ref{en2}), we
get the reduced expression of the normalized entanglement
negativity as
\begin{eqnarray}
N^{N}(\rho_{M})= \frac{54 N^{D}(\rho_{M})}{9153}\left[169+\frac{N^{D}(\rho_{M})}{2}\right]. \label{enex1}
\end{eqnarray}
When we compare the expressions given in (\ref{entnegdef1}) and
(\ref{enex1}) then we find that the two curves for
$N^{D}(\rho_{M})$ and $N^{N}(\rho_{M})$ almost coincide with each other
and it can be verified by Fig. 1 also.

\begin{figure}[h]
\centering
\includegraphics[scale=0.6]{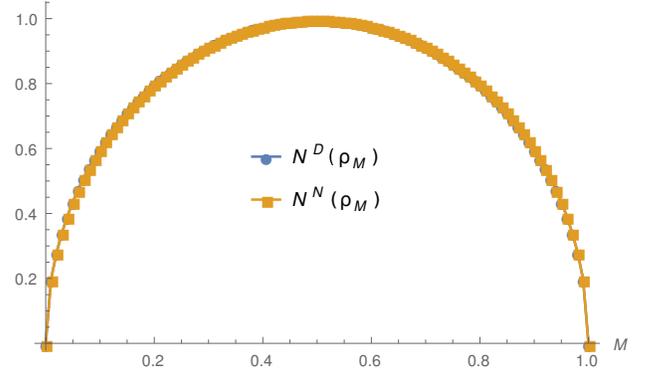}
\caption{Plot of negativity obtained using definition given by
$N^{D}(\rho_{M})$ and obtained by our formula $N^{N}(\rho_{M})$.
}
\end{figure}

\textbf{Example-2:} Next, consider a mixed entangled state also
known as Horodecki state given by \cite{mhorodecki}
\begin{eqnarray}
\rho_{H}=p|\psi^{+}\rangle\langle\psi^{+}|+(1-p)|00\rangle\langle00|
\label{mixedent}
\end{eqnarray}
The entanglement negativity using the definition (\ref{def}) is
given by \cite{miranowicz1}
\begin{eqnarray}
N^{D}(\rho_{H})=\sqrt{(1-p)^{2}+p^{2}}-(1-p) \label{entnegdef2}
\end{eqnarray}
The SPA-PT of $\rho_{H}$ is given by
\begin{eqnarray}
\widetilde{\rho_{H}}=
\begin{pmatrix}
  \frac{3-p}{9} & 0 & 0 & \frac{p}{18} \\
  0 & \frac{2+\frac{p}{2}}{9} & \frac{ip}{18} & 0 \\
  0 & \frac{-ip}{18} & \frac{2+\frac{p}{2}}{9} & 0 \\
   \frac{p}{18}  & 0 & 0 & \frac{2}{9}
\end{pmatrix}
\label{spapt2}
\end{eqnarray}
The minimum eigenvalue of $\widetilde{\rho_{H}}$ is given by
\begin{eqnarray}
\mu_{min}&=&\frac{5}{18}-\frac{p}{18}-\frac{\sqrt{1-2p+2p^{2}}}{18}\nonumber\\&=&
\frac{15}{8}F_{avg}(\widetilde{\textit{\textbf{W}}},\widetilde{\rho_{H}})-\frac{47}{72}
 \label{mineig2}
\end{eqnarray}
In this example also, we can follow the same procedure as
explained in previous example to estimate the minimum eigenvalue
experimentally. Using (\ref{mineig2}) in (\ref{en2}), the
expression of the normalized entanglement negativity reduced as
\begin{eqnarray}
N^{N}(\rho_{H})&=&
\frac{N^{D}(\rho_{H})}{339}\left[338+N^{D}(\rho_{H})\right].
\label{enex2}
\end{eqnarray}
Again if we compare the expressions given in (\ref{entnegdef2})
and (\ref{enex2}) then we can see that the two curves for
$N^{D}(\rho_{H})$ and $N^{N}(\rho_{H})$ almost overlap with each other.

\begin{figure}[h]
\centering
\includegraphics[scale=0.9]{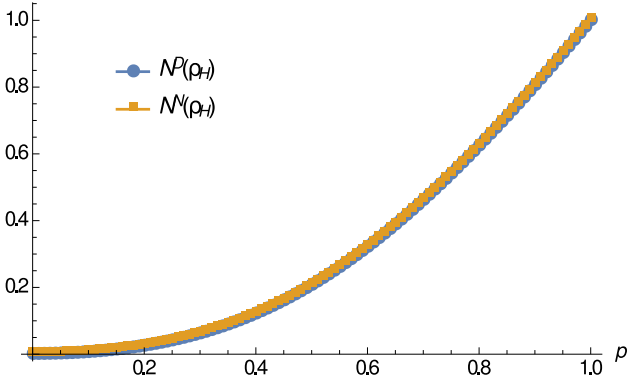}
\caption{Plot of negativity obtained using definition given by
$N^{D}(\rho_{H})$ and obtained by our formula $N^{N}(\rho_{H})$.
}
\end{figure}

\section{Determination of exact value of concurrence}
The first entanglement measure popularly known as concurrence was
introduced by Wootters \cite{wootters} and it is defined for the
two-qubit state $\rho$ as
\begin{eqnarray}
C(\rho)=max\{0,\lambda_{1}-\lambda_{2}-\lambda_{3}-\lambda_{4}\}
 \label{con}
\end{eqnarray}
where $\lambda_{1},\lambda_{2},\lambda_{3},\lambda_{4}$ denote the
square root of the eigenvalues of the operator
$\rho(\sigma_{y}\otimes\sigma_{y})\rho^{*}(\sigma_{y}\otimes\sigma_{y})$
and these eigenvalues are arranged in descending order,
$\sigma_{y}$ represent the Pauli spin matrix and complex
conjugation is denoted by asterisk.\\
We find that it is not possible to express always the exact value
of concurrence in terms of a parameter that can be realized
experimentally. In this section, we express the exact formula of
the concurrence in terms of experimentally accessible parameter
for two classes of entangled states, viz., for pure entangled
states and for quasi-distillable mixed entangled states.
\subsection{Pure entangled states}
It has been shown that the entanglement negativity and concurrence
are equal for pure two-qubit state \cite{audenaert}. Therefore,
the concurrence for any pure two-qubit state $|\psi\rangle_{AB}$
can be expressed in terms of the minimum eigenvalue $\mu_{min}$ of
$\widetilde{\rho_{AB}}$, which is a SPA-PT of a two-qubit state
$|\psi\rangle_{AB}$. Thus, the concurrence for the pure state
$|\psi\rangle_{AB}$ is given by
\begin{eqnarray}
C(|\psi\rangle_{AB})&=&
\frac{108}{113}(\frac{2}{9}-\mu_{min})(19-\mu_{min}),\nonumber\\&&
\frac{1}{6}\leq \mu_{min} < \frac{2}{9} \label{con1}
\end{eqnarray}
Therefore, the value of the concurrence for any two-qubit pure
state can be realized experimentally.
\subsection{Quasi-distillable mixed states}
Quasi-distillable states are those non-maximally entangled states
which cannot be distilled to a perfect singlet state but can be
distilled to a state with arbitrarily high singlet fraction and
this distillation procedure can be implemented with non-zero
probability \cite{horodecki}. The class of quasi-distillable
states are given by
\begin{eqnarray}
\rho_{AB}^{quasi}=
\begin{pmatrix}
  \frac{C}{2} & 0 & 0 & \frac{C}{2} \\
  0 & 1-C & 0 & 0 \\
  0 & 0 & 0 & 0 \\
  \frac{C}{2} & 0 & 0 & \frac{C}{2}
\end{pmatrix}
\end{eqnarray}
where $C$ denotes the concurrence of the state
$\rho_{AB}^{quasi}$.\\
Verstraete et. al. \cite{verstraete} have shown that for any
two-qubit entangled state $\rho_{AB}$, entanglement negativity
$N(\rho_{AB})$ and concurrence $C(\rho_{AB})$ are related by the
inequality
\begin{eqnarray}
N(\rho_{AB})& \geq &
\sqrt{(1-C(\rho_{AB}))^{2}+C(\rho_{AB})^{2}}\nonumber\\&&
-(1-C(\rho_{AB}))
 \label{negconineq}
\end{eqnarray}
The lower bound of (\ref{negconineq}) is achieved iff the state is
a rank-2 quasi-distillable state \cite{verstraete}. Therefore, the
entanglement negativity and concurrence for rank-2
quasi-distillable state $\rho_{AB}^{quasi}$ are related as
\begin{eqnarray}
N(\rho_{AB}^{quasi})&=&
\sqrt{(1-C(\rho_{AB}^{quasi}))^{2}+C(\rho_{AB}^{quasi})^{2}}\nonumber\\&&
-(1-C(\rho_{AB}^{quasi}))
 \label{negconeq}
\end{eqnarray}
We note that the entanglement negativity appeared in equation
(\ref{negconeq}) is normalized and hence the concurrence for
rank-2 quasi-distillable state $\rho_{AB}^{quasi}$ is explicitly
espressed in terms of normalized entanglement negativity as
\begin{eqnarray}
C(\rho_{AB}^{quasi})=-N(\rho_{AB}^{quasi})+\sqrt{2N(\rho_{AB}^{quasi})(N(\rho_{AB}^{quasi}+1)}
\label{connegeq}
\end{eqnarray}
where $N(\rho_{AB}^{quasi})$ is given by equation (\ref{en2}).
Since the value of $N(\rho_{AB}^{quasi})$ can be estimated
experimentally so the concurrence for rank-2 quasi-distillable
state can be estimated experimentally.
\section{Upper bound of the concurrence for the general two-qubit mixed
states} Numerous effort had been given to calculate the lower and
upper bound of concurrence theoretically or experimentally
\cite{mintert1, mintert2, mintert3, jurkowski, augusiak}. All
these methods rely on the joint measurement on two copies of the
given quantum state. In this section, we will show that the upper
bound of the concurrence can be estimated using only single copy
of the given quantum state.\\
We now derive the upper bound of the concurrence using
Lewenstein-Sanpera decomposition \cite{lewenstein}  and the
convexity property of concurrence. An arbitrary two-qubit density
matrix $\rho$ can be decomposed as \cite{lewenstein}
\begin{eqnarray}
\rho=\lambda \rho_{s}+(1-\lambda)|\psi\rangle_{e}\langle\psi|,
\lambda \in [0,1] \label{l-sdecom}
\end{eqnarray}
where $\rho_{s}$ denote two-qubit separable state and
$|\psi\rangle_{e}$ represent two-qubit pure entangled state. We
note here that the decomposition (\ref{l-sdecom}) is unique.\\
The concurrence of an arbitrary two-qubit quantum state $\rho$ is
given by
\begin{eqnarray}
C(\rho)=C(\lambda
\rho_{s}+(1-\lambda)|\psi\rangle_{e}\langle\psi|)
\label{concl-sdecom}
\end{eqnarray}
Using the convexity property of concurrence \cite{uhlmann1,
vedral}, we have
\begin{eqnarray}
C(\rho)&\leq&
(1-\lambda)C(|\psi\rangle_{e}\langle\psi|))\nonumber\\&=&
\frac{108}{113}(1-\lambda)(\frac{2}{9}-\mu_{min})(19-\mu_{min}),\nonumber\\&&\frac{1}{6}\leq
\mu_{min} < \frac{2}{9} \label{conclowbound}
\end{eqnarray}
In the first line, we have used $C(\rho_{s})=0$ and equation
(\ref{con1}) is used in the second line. The equality holds when
$\lambda=0$ and $\lambda=1$.
\section{Conclusion} To summarize, in this work we have studied the two most important measure of
entanglement such as entanglement negativity and concurrence to
quantify the entanglement in a two-qubit system. As partial
transposition (PT) of a two-qubit density matrix is not a physical
operation so we apply structural physical approximation (SPA)
method to realize the partial transposition operation
experimentally. It has already been shown that the minimum
eigenvalue of SPA-PT state can be estimated experimentally
\cite{adhikari}. Interestingly, we show that there is a connection
between the entanglement negativity and minimum eigenvalue of
SPA-PT state and hence entanglement negativity can be estimated
experimentally. We have provided some illustrations to show that
the value of the entanglement negativity would be obtained
experimentally matches with the theoretical result. Further, we
have shown that the value of the concurrence can be estimated
experimentally for (i) any arbitrary two-qubit pure states and
(ii) a particular class of mixed state known as rank-2
quasi-distillable states. Also we have obtained the upper bound of
the concurrence of an arbitrary two-qubit quantum state. We can
estimate the value of entanglement negativity and concurrence
using Hong-Ou-Mandel set up that require only two measurements,
which are much lesser in comparison to state tomography of an
unknown state, thus signifying the practical utility of our work.
Since we restrict ourselves in this work only to two-qubit system
so it would be interesting to apply the idea of this work to
obtain the experimental value of entanglement negativity and
concurrence for higher dimensional system or for multi-partite
system.\\

\textit{Acknowledgement:} I would like to thank Shuming Cheng for
pointing out the error in Figure-1 in the earlier version. I would also like to thank
Sk. Sazim and Chandan Dutta for the plot of figures.

\end{document}